\documentclass[doublecol]{epl2}

\usepackage{graphics}
\usepackage[T1]{fontenc}
\usepackage[latin1]{inputenc}
\usepackage{amsmath}
\usepackage{graphicx}
\usepackage{amssymb}

\usepackage{graphicx}
\usepackage{epsfig}
\usepackage{multirow}
\usepackage{dcolumn}
\usepackage{bm}
\usepackage{subfigure}

\def\<{\langle}
\def\>{\rangle}

\newcommand{\cceb}{u_{\mbox{\small e-b}}}
\newcommand{\ccbb}{u_{\mbox{\small b-b}}}

\newcommand{\leads}{\mbox{\small leads}}
\newcommand{\edge}{e}
\newcommand{\teins}{$|t_1\rangle$}
\newcommand{\tdrei}{$|t_3\rangle$}
\newcommand{\geins}{$|g_1\rangle$}
\newcommand{\gzwei}{$|g_2\rangle$}
\newcommand{\tzwei}{$|t_2\rangle$}
\newcommand{\ezwei}{$|e_2\rangle$}

\begin{document}

\title{Graphene armchair nanoribbon single-electron transistors: The peculiar influence of end states}
\author{Sonja Koller\inst{1}, Leonhard Mayrhofer\inst{1,2}, and Milena Grifoni\inst{1}}
\institute{
\inst{1}Theoretische Physik, Universit\"at Regensburg, 93040 Germany\\
\inst{2}Fraunhofer IWM Freiburg, Wöhlerstraße 11, 79108 Freiburg
}

\date{\today}

%
%
\abstract{
We present a microscopic theory for interacting graphene armchair nanoribbon quantum dots.
Long range interaction processes are responsible for Coulomb blockade and spin-charge separation.
Short range ones, arising from the underlying honeycomb lattice of graphene smear the spin-charge separation and induce exchange
correlations between bulk electrons -- delocalized on the ribbon -- and single electrons localized at the two ends.
As a consequence, entangled end-bulk states where the bulk spin is no longer a conserved quantity occur.
Entanglement´s signature is the occurrence of negative differential conductance effects in a fully symmetric set-up due to symmetry-forbidden transitions.
}
%

\maketitle
%
%
%
The first successful separation of graphene \cite{Novoselov04}, a single atomic layer of graphite, has resulted in intense theoretical and experimental investigations on graphene-based structures \cite{Castro-Neto08}, because of potential applications and fundamental physics issues  arising from the linear dispersion relation
in the electronic band structure of graphene.\\
In graphene nanostructures, confinement effects typical of mesoscopic
systems and electron-electron interactions are expected to play a crucial
role on the transport properties. Indeed a tunable single-electron
transistor has been demonstrated in a graphene island weakly coupled to
leads \cite{Stampfer08}. Conductance quantization has been observed in 30nm
wide ribbons \cite{Lin08}, while an energy gap near the charge neutrality
point scaling with the inverse ribbon width was reported in \cite{Han07}.
Theoretical investigations \cite{Sols07,Zarea07} have attributed the
existence of such a gap to Coulomb interaction effects.\\
Confinement is also known to induce localized states at zig-zag boundaries \cite{wakabayashi96},
possessing a flat energy band and occuring in the mid of the gap.
Those states have been analysed \cite{Wunsch08} under the assumption of a filled valence and an empty conduction
band (half-filling), taking into account both Hubbard and long-ranged Coulomb interaction.
There was a prediction of strong spin features in case of a low population of these midgap states. 

Above the half-filling regime, however, no detailed study on the interplay between
longitudinal quantization effects and Coulomb interactions in the spectrum
of narrow nanoribbons exists at present.
\begin{figure}
\begin{center}
\includegraphics[width=0.85\columnwidth]{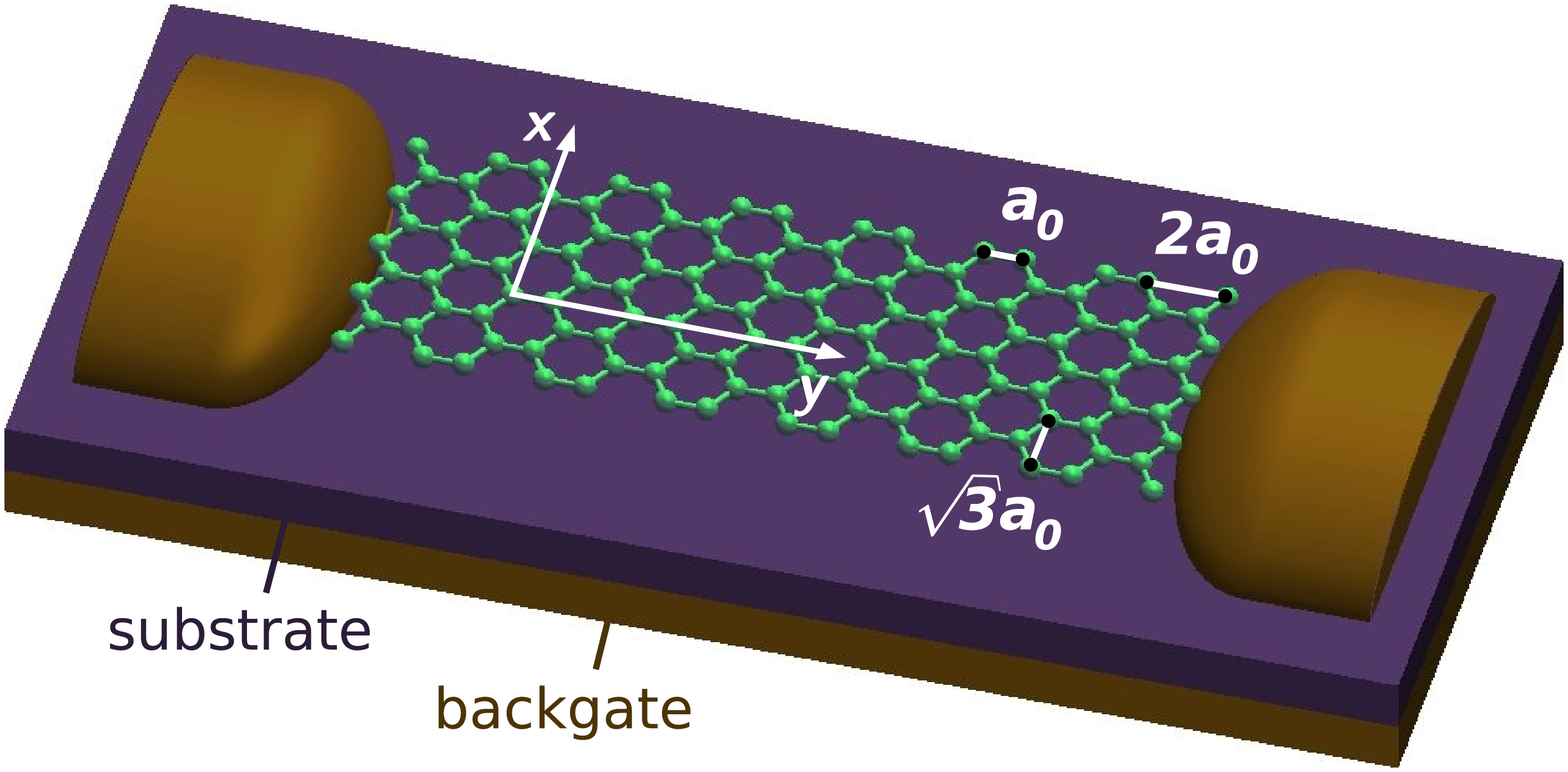}
\end{center}
\caption{\label{graphene}A graphene armchair nanoribbon single-electron transistor. At the long sides, the lattice is terminated in armchair, at the small ends in zig-zag configuration.
}
\end{figure}

The purpose of this Letter is to derive a low energy theory of armchair nanoribbons (ACN) single-electron transistors (SETs),
  see Fig. \ref{graphene}, i.e., to investigate the consequences of confinement and
  interaction in narrow ACNs weakly coupled to leads. Short ACN have recently been synthesized \cite{Yang08}.
We show that the long-range part of the Coulomb interaction is responsible for charging effects and spin-charge separation.
Short-range processes, arising due to the presence of two atoms per unit cell in graphene as well as of localized end states, lead to exchange coupling.
Bulk-bulk short-range interactions have only a minor effect on the energy spectrum. However, interactions between end states localized at the  narrow zig-zag ends of the stripe and bulk states smear the spin-charge separation. Moreover, they cause an entanglement of end-bulk states with the same total spin. Hence, despite the weak spin-orbit coupling, the bulk spin is not  a conserved quantity in ACNs. These  states strongly influence the nonlinear transport.  We predict the occurrence of   negative differential conductance (NDC), due to symmetry-forbidden transitions between entangled states,   in a fully symmetric setup.\\
We proceed as follows: in the first part of this Letter we set up the interacting Hamiltonian of ACNs and derive their energy spectrum.
In a second part transport in the single electron tunneling regime is investigated.

\section{Electron operator of a metallic ACN}
\begin{figure}
\includegraphics[width=0.99\columnwidth]{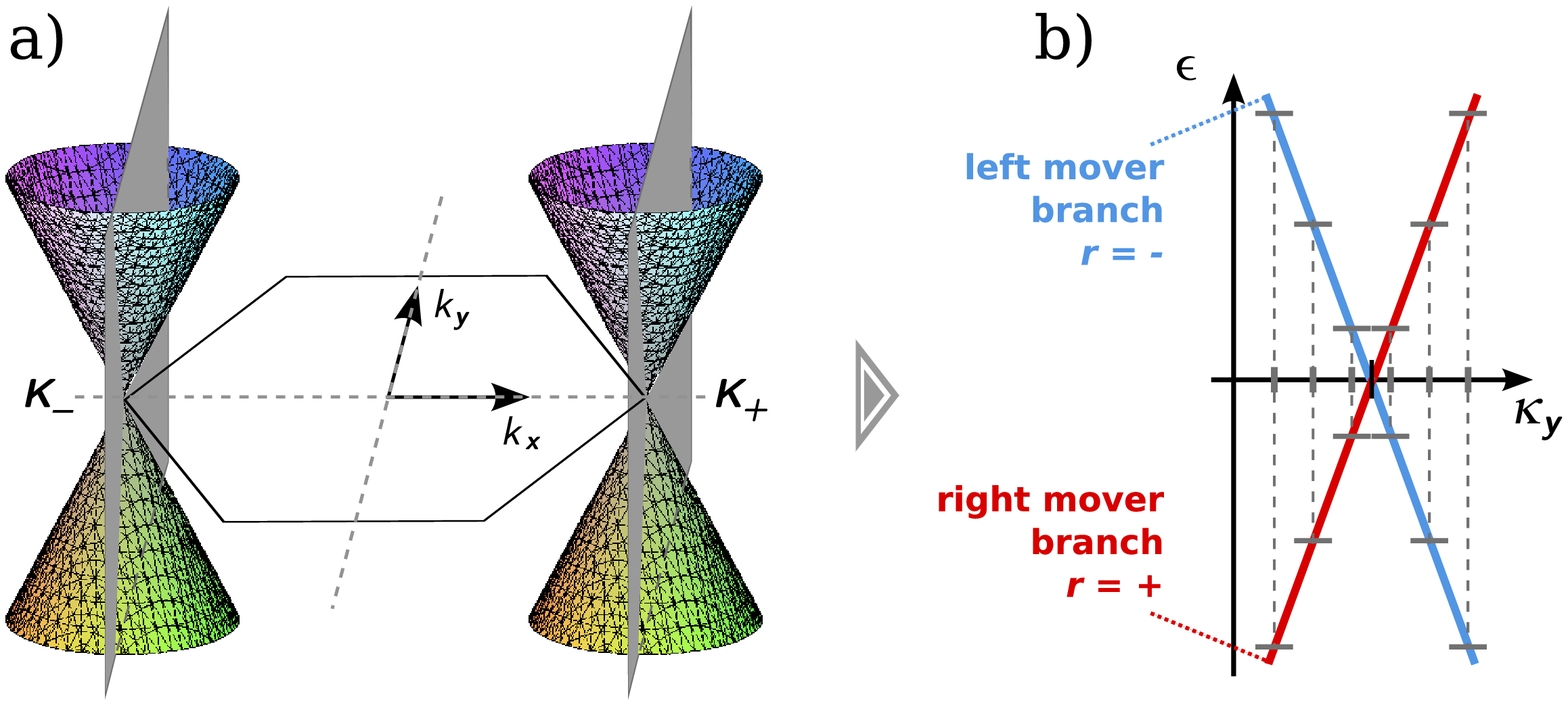}
\caption{\label{disp1}a) Dispersion relation of a graphene stripe for real momenta.  In the low energy regime, only subbands lying on the plane  $\kappa_x=0$ play a role due to the condition $L_x\ll L_y$, with $\vec\kappa$ the vector relative to the Dirac point $K_+$ $(K_-)$. b) Confinement along the ACN length yields  quantization of $\kappa_y$. }
\end{figure}

The carbon atoms in graphene are  arranged in a  honeycomb
lattice. There are two atoms per unit cell  that  define
two different sublattices $p=\pm$.
Overlapping
 $2p_{z}$ orbitals form  valence and conduction $\pi$-bands that touch
at the corner points of the first Brillouin zone, also called Dirac
points, and determine the
electronic properties at low energies.
From now on we focus on the region of linear dispersion in the vicinity of the two inequivalent Dirac points, see  Fig. \ref{disp1}a, $K_{F}=F\frac{4\pi}{3\sqrt{3}a_{0}}\hat{e}_{x},\, F=\pm, $
where $a_{0}\approx0.14\mbox{ nm}$ is the nearest neighbour distance.
%
%
%
%
%
Then  the $\pi$-electrons are described by Bloch waves
\begin{eqnarray}
\varphi_{F\alpha}(\vec{r},\vec{\kappa})&=&\frac{1}{\sqrt{{2N}_{L}}}
\sum_{p=\pm}\eta_{F\alpha p}(\vec{\kappa})\, \sum_{\vec{R}}e^{i\left(\vec{K}_{F}+\vec{\kappa}\right) \cdot\vec{R}}\chi_{\vec{R}\, p}(\vec{r})\nonumber\\&=:&\sum_{p=\pm}\eta_{F\alpha p}(\vec{\kappa})\, \varphi_{Fp}(\vec{r},\vec{\kappa}),\label{eq:bloch waves}\end{eqnarray}
 where $N_{L}$ is the number of sites of the considered lattice,
$\alpha=\pm$ denotes the conduction/valence band,
and $\chi_{\vec{R}\, p}(\vec{r})$ is the $2p_{z}$ orbital on sublattice
$p$ at lattice site $\vec{R}$, with $\vec{r},\,\vec{R}\in\mathbb{R}^2$. Furthermore $\vec{\kappa}=(\kappa_x,\kappa_y)$ is the wave
vector relative to the Dirac point $\vec{K}_{F}$. Finally, the spinors
$\eta_{F\alpha}(\vec{\kappa}):=(\eta_{F\alpha -}(\vec{\kappa}),\eta_{F\alpha +}(\vec{\kappa}))$
  fulfill the Dirac equation with a velocity $v_F=8.1 \cdot 10^5\,$m/s.
%

To describe ACNs  boundary conditions have to be assumed.
 Following Ref. \cite{Brey06}  we demand that the wavefunction  vanishes
  on sublattice $p=-$ on the left end, $y=0$,  and
on $p=+$ on the right end, $y=L$.
At the armchair edges the terminating
atoms where the wave function is required to vanish are from both
sublattices.
The quantization condition from the zigzag ends reads \cite{Brey06}
 %
  \begin{equation}
e^{i2\kappa_{y}L_{y}}=
\left(F\kappa_{x}+i\kappa_{y}\right)/\left(F\kappa_{x}-i\kappa_{y}\right);
\label{eq:zigzagquantcondition}\end{equation}
that from the armchair edges is
$ K_{+}+\kappa_{x}=\frac{\pi}{L_{x}}n_{x},$ $n_{x}\in\mathbb{Z}.$
Eq. (\ref{eq:zigzagquantcondition})  supports the presence of extended states -- real $\kappa_y $ -- as well  of localized states -- purely imaginary $\kappa_{y}$ \cite{wakabayashi96}.

 Let us first discuss the bulk states.
Due to $L_{y}\gg L_{x}$ the longitudinal quantization condition
yields subbands assigned to different $\kappa_{x}.$
From now on we focus on the low energy regime
of metallic ACNs, where only the gapless subbands [$\kappa_{x}=0$, Fig. \ref{disp1}a)] are relevant. Eq. (\ref{eq:zigzagquantcondition}) yields then
$ \kappa_{y}=(n_y+\frac{1}{2})\frac{\pi}{L_{y}},\, n_y\in\mathbb{Z}$, Fig. \ref{disp1}b).
%
%
Bearing in mind
Eq. (\ref{eq:bloch waves}),
we can finally express the states $\varphi_{\kappa_{y}}$ in terms
of the sublattice wave functions $\varphi_{Fp}$,\begin{equation*}
\varphi^{\mbox{\small }}_{\kappa_{y}}(\vec{r})=\frac{1}{2}\sum_{Fpr=\pm}Ff_{pr}\,\varphi_{Fp}(\vec{r},(0,\kappa_{y})),
\end{equation*}
 where $r=\pm$ denotes right/left moving waves. Up to a complex prefactor, the coefficients are $f_{+r}=r$, $f_{-r}=i$.

The quantization condition (\ref{eq:zigzagquantcondition})
also allows purely imaginary $\kappa_{y}$:
  For each $\kappa_{x}=n_{x}\pi/L_{x}>1/L_{y},\, n_{x}\in\mathbb{N}$ there exist two imaginary solutions
  $\kappa_{y}(\kappa_{x})$.
  Besides, due to $L_{x}\ll L_{y}$, it holds to a very good
approximation  $
\kappa_{y}(\kappa_{x})=\pm i\kappa_{x}$.
 The corresponding ACN eigenstates can be chosen to live
on one sublattice $p=\pm$ only: \begin{equation*}
\varphi_{p \kappa_{x}}^{\mbox{\small e}}(\vec{r})=C(\kappa_x)\sum_{F}F\varphi_{Fp}(\vec{r},(F\kappa_{x},ip\kappa_x)),
\end{equation*}where $C(\kappa_{x})$ is a normalization constant.
 The decay length
of $\varphi_{p\kappa_{x}}^{\mbox{\small e}}$ from one of the zigzag ends to
the interior
is $\kappa_x^{-1},$
which is
 much shorter than the ribbon length. Hence end states are localized.
 %
%
  From the graphene dispersion relation
 it follows that the energy of the
end states is zero. They will be unpopulated below half filling
but as soon as the Dirac point is reached one electron will get trapped at each end.
 For small width ribbons the strongly localized character of the end states implies
  Coulomb addition energies for a second electron on the same end
 by far exceeding the addition energy for the bulk states.
  Thus at  low energies above the Dirac points  both end states are populated
  with \emph{a single electron only}. Introducing bulk and end electron annihilation operators
  $c_{\sigma\kappa_y }$, $d_{\sigma p\kappa_x }$,
  the noninteracting Hamiltonian is
  \begin{equation}H_{0}=\hbar v_F\sum_{\sigma\kappa_{y}}\kappa_{y}c_{{\sigma\kappa}_{y}}^{\dagger}
c_{{\sigma\kappa}_{y}},\label{H0}\end{equation} because the end states have zero energy, and the field
operator for an electron with spin $\sigma$ at position $\vec{r}$
is \begin{equation}
\Psi_{\sigma}(\vec{r})=\underbrace{\sum_{\kappa_{y}}\varphi^{\mbox{\small }}_{\kappa_{y}}(\vec{r})c_{\sigma \kappa_{y}}}_{=:\psi^{\mbox{\small }}_{\sigma}(\vec{r})}+\sum_{p}\underbrace{\sum_{\kappa_{x}}\varphi_{p\kappa_{x}}^{\mbox{\small e}}(\vec{r})
d_{\sigma p\kappa_{x}}}_{=:\psi^{\mbox{\small e}}_{p\sigma}(\vec{r})}.
\label{elops}\end{equation}
 The 1D character of ACNs at low energies becomes evident by defining
the slowly varying electron operators 
$\psi_{r\sigma}(y):=\frac{1}{\sqrt{2L_y}}\sum_{\kappa_{y}}e^{ir \kappa_y y}c_{\sigma \kappa_{y}}$ 
 such that we obtain
 \begin{equation}
\psi_{\sigma}(\vec{r})=\sqrt{L_y/2}
\sum_{pr}Ff_{pr}\varphi_{Fp}(\vec{r})\psi_{r\sigma}(y),\label{eq:3Dto1D}\end{equation}
where $\varphi_{Fp}(\vec{r})\!:=\!\varphi_{Fp}(\vec{r},\vec{\kappa}\!=\!\vec{0})$. 
%
\section{Hamilton operator of the interacting ACN}
Including the relevant Coulomb interactions yields the total Hamiltonian
\begin{equation}H_\odot=H_0+V_{\mbox{\small e-b}}+V_{\mbox{\small b-b}}.
\label{eq:H_dot}
\end{equation}
First, there is interaction between end and bulk states,
\begin{equation*}
{V}_{\mbox{\small e-b}}=\frac{L_y}{2}\sum_{\kappa_x}\sum_{\sigma\sigma' rr' p}\psi_{r\sigma}^{\dagger}(y_p)\psi_{r'\sigma'}(y_p)u^{\kappa_x}_{\mbox{\small e-b}}d_{\sigma'p\kappa_x}^{\dagger}d_{\sigma p\kappa_x}\,,\label{V_e-b}
\end{equation*}with $y_{\pm}=0/L$ and with $U_{3D}(\vec{r}-\vec{r}\,')$ denoting the 3D Coulomb potential, the coupling constant
\begin{align}
\cceb^{\kappa_x}=&\sum_{FF'}FF'\int\!\!\!\!\int d\vec{r} d\vec{r}\,'\varphi_{F+}^{*}(\vec{r})\varphi_{F'+}(\vec{r})U_{3D}(\vec{r}-\vec{r}\,')\nonumber\\
&\times \varphi_{+\kappa_x}^{\mbox{\small \edge}*}(\vec{r}\,')\varphi_{+\kappa_x}^{\mbox{\small \edge}}(\vec{r}\,').\label{t}
\end{align}For ACNs of width $L_x$ ranging from $5$ to $25$\,nm, one finds from numerical evaluation $\cceb^{\kappa_x} \approx \cceb$, with $\cceb L_x/\varepsilon_0 \approx 0.55\,$nm, practically independent of $\kappa_x$.

Secondly, interaction between the extended bulk states,
\begin{equation*}
\hspace{-2mm}V_{\mbox{\small b-b}}=\sum_{S_r=u,b,f^{\pm}}\sum_{S_\sigma=f^{\pm}}V^{\mbox{\small b-b}}_{S_rS_\sigma},
\end{equation*}
\begin{figure}
\begin{center}\includegraphics[width=0.7\columnwidth]{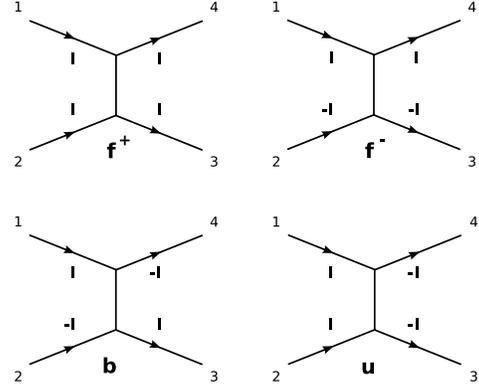}\end{center}
\caption{\label{scattering}The four different possibilities for scattering are forward \mbox{($f^{\pm}$)-,} back ($b$)-, and umklapp ($u$)- processes.
As it can be seen from the sketch, those correspond to different relations between a certain index $I$ of the states before and after the scattering event.
 }
\end{figure}

%
\noindent is classified by the scattering types $S_r,S_\sigma$ concerning band and spin, respectively, where one distinguishes between forward \mbox{($f^{\pm}$)-,} back ($b$)-, and umklapp ($u$)- scattering.
Denoting the scattering type by $S_{I}$ we define $[I]_{S_{I}=f^{\pm}}:=[I,\pm I,\pm I,I]$, $[I]_{b}:=[I,-I,I,-I]$ and $[I]_{u}:=[I,I,-I,-I]$, see also Fig. \ref{scattering}.
With Eq. (\ref{eq:3Dto1D}) one finds
\begin{multline}
  V^{\mbox{\small b-b}}_{S_{r}S_{\sigma}}=\frac{1}{2}\sum\nolimits_{\{[r]_{S_{r}},[\sigma]_{S_{\sigma}}\}}\int\!\!\int dydy'\times\\\times\psi_{r_{1}\sigma}^{\dagger}(y)\psi_{r_{2}\sigma'}^{\dagger}(y')U_{[r]_{S_r}}(y,y')
  \psi_{r_{3}\sigma'}(y')\psi_{r_{4}\sigma}(y). \label{V_bb_rs}\end{multline}
Hereby, the potential mediating the interactions is either
\begin{equation*}
U_{[r]_{f^\pm}}=U^{\mbox{\footnotesize intra}}+U^{\mbox{\footnotesize inter}}\quad\mathrm{or}\quad\ U_{[r]_{b,u}}=U^{\mbox{\small intra}}- U^{\mbox{\footnotesize inter}},
\end{equation*} where the 1D potentials $U^{\mbox{\footnotesize intra/inter}}$ describe interactions between electrons on the same/different sublattice \cite{Mayrhofer2008}. 
While end-bulk scattering is completely short-ranged, the bulk-bulk interactions split into long-/short-ranged contributions ($S_r$=$f^\pm/S_r$=$u,b$).
The short-range bulk-bulk coupling constant is
\begin{equation}
\ccbb=\frac{1}{4L_y^2}\int\!\!\!\!\int dydy' U_{[r]_{b,u}}(y,y').\label{u}
\end{equation}
  %
  %
The long-ranged part of the interaction is diagonalizable
 by bosonization \cite{Delft1998}. We find
\begin{equation}
H_{0}+V^{\mbox{\small b-b}}_{\mbox{\small long}}=
\frac{1}{2}W_{0} N_{c}^{2}
+\frac{1}{2}\varepsilon_{0}\sum_\sigma (N_\sigma+N^2_{\sigma}) +H_{\mbox{\small bos}}.
\label{eq:H0Vrr_diag}\end{equation}
The first term of (\ref{eq:H0Vrr_diag}), with $N_{c}=\sum_{\sigma} N_{\sigma}$ being the charge operator on the ACN,
 $W_0=W_{q=0}$ with
 \begin{equation*}
W_{q}=\frac{1}{2L_y^{2}}\int\!\!\int dydy'U_{[r]_{f^\pm}}(y,y')\cos(qy)\cos(qy'),
\end{equation*}
 accounts for Coulomb charging effects. The second  term, where $\varepsilon_0=\hbar v/L$ is the level spacing, yields the fulfillment of Pauli exclusion principle. Finally, $H_{\mbox{\small bos}}=
\sum_{j,q>0}\varepsilon_{j q}a_{j q}^{\dagger}a_{j q}$ accounts for the bosonic excitations of the system, created/annihilated
by the operators $a_{j q}^{\dagger}$ / $a_{j q}$. The two
 channels $j=c,s$ are associated to charge $(c)$ and spin $(s)$ excitations.
The excitation energies are $\varepsilon_{sq}=n_q\varepsilon_0, \;\;\varepsilon_{cq}=n_q\varepsilon_{0}\sqrt{1+2W_q/\varepsilon_0}$ with $n_q\in \mathbb{N}$.\\
%
Eigenstates of $H_{0}+V^{\mbox{\small b-b}}_{\mbox{\small long}}$ are
  $|{\sigma}_L^{\mbox{\small \edge}},\vec{N},\vec{m},{\sigma}_R^{\mbox{\small \edge}}\rangle $,
  where $\vec{m}$ characterizes the bosonic excitations, and the fermionic configuration $\vec{N}=(N_{\uparrow},N_{\downarrow})$ defines the number of electrons in each spin band.
 Above half filling exactly one electron occupies each end state and thus the end configurations  $\sigma^{\mbox{\small \edge}}_L,\sigma^{\mbox{\small \edge}}_R\in\{\uparrow,\downarrow\}$. 

These states can be used
as basis to examine the effect of $V_{\mbox{\small b-b}}$ and $V_{\mbox{\small e-b}}$ on the spectrum of an interacting ACN.
For this purpose one needs to evaluate  the corresponding matrix elements
 proportional to the short-range coupling constants $\cceb$, Eq.~(\ref{t}),
 and $\ccbb$, Eq.~(\ref{u}). As the procedure follows similar lines as in \cite{Mayrhofer2008} we refrain 
 from reporting it here and discuss the main results.\\
 A diagonalization of the full Hamiltonian yields energy spectrum
 and eigenstates of the system including both long and short-range
 interactions. As those are spin preserving, it is clear that linear
 combinations must be formed of states with same spin-$S_z$ component.
 Thereby, importantly, the end spin degrees of freedom permit a mixture
 between states of different bulk spin configurations. This mechanism and
 its impacts will be illuminated in the course of the following sections.
 \section{Spectrum of interacting ACNs}
\begin{figure}
\begin{center}\includegraphics[width=0.99\columnwidth]{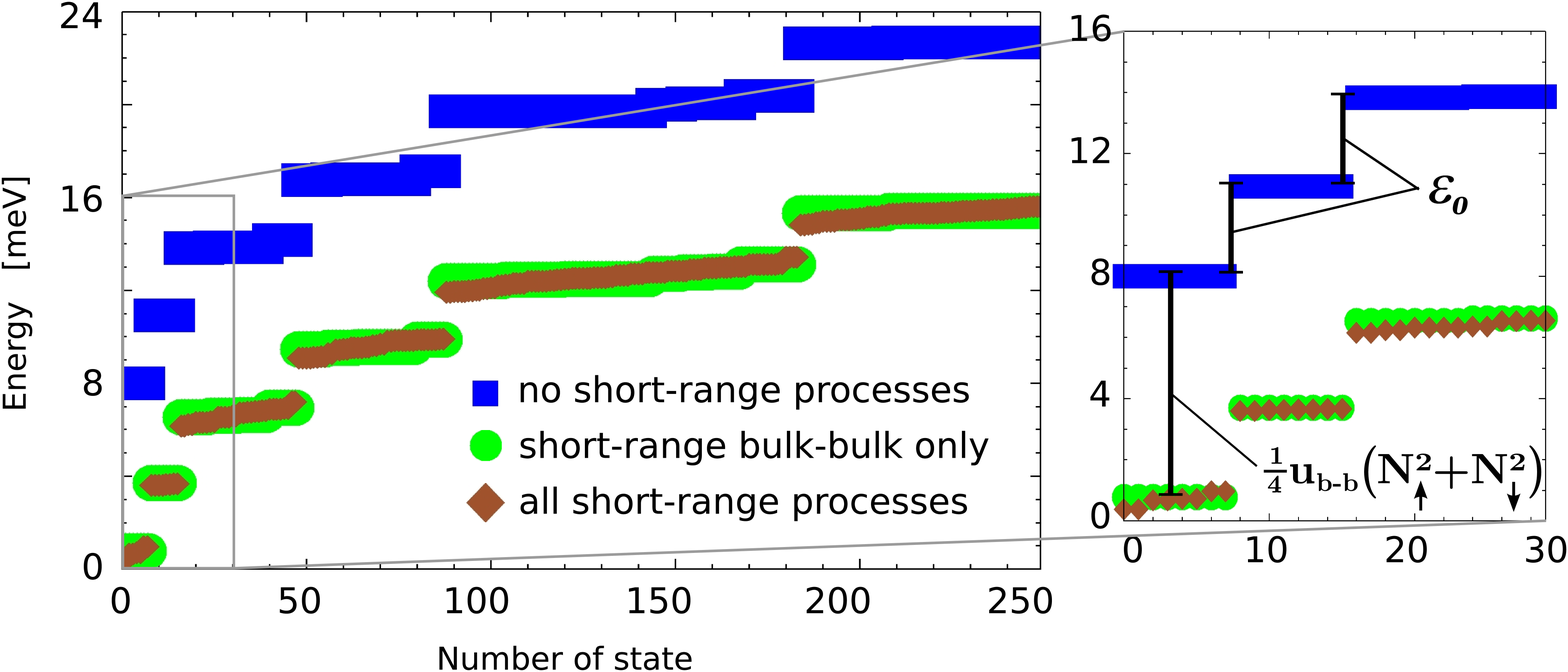}\end{center}
\caption{\label{spectrum}The spectrum of an ACN with $N_c=2n+1$ electrons.
We chose a $7.8\,$nm width and $572\,$nm length, corresponding to a
charging energy $W_0=2.3\,$meV, and to short-range bulk-bulk and end-bulk
coupling constants $\ccbb=0.036\,$meV, $\cceb=0.21\,$meV. End-bulk scattering i) mixes end and bulk states,  ii) spoils the spin-charge separation. The inset zooms on the lowest lying 30 states.
 }
\end{figure}
Numerical calculation and diagonaliziton of the full ACN Hamiltonian including the 250 lowest lying states of a $572\,$nm $\times$ $7.8\,$nm ribbon leads to the spectrum found in Fig. \ref{spectrum}. For comparison we also give the energies without the end-bulk interaction and for long-range interactions only. From Eq. (\ref{eq:H0Vrr_diag}) it can be found that without short-range interactions (blue squares), the energy cost for both a fermionic and a bosonic spin-like excitation amounts to $\varepsilon_0$. That is why in the spectrum discrete plateaus which are separated by this energy arise. The first charge-like bosonic mode can be excited at an energy of about $2.1\varepsilon_0$, which shows up in form of a small step towards the end of the third and all following plateaus. Switching on the short-range bulk-bulk contributions (green disks) actually preserves this \emph{spin-charge separation}: while the curve as a whole is shifted downwards in energy due to an exchange term (see inset of Fig. \ref{spectrum}), all steps within the plateaus remain resolvable. In contrast to what is found for carbon nanotubes~\cite{Mayrhofer2008}, there is only a very tiny additional lifting due to the bulk-bulk exchange, which cannot compare in magnitude with the spin-charge separation. The deeper reason is that, as it can be seen from an explicit calculation, only the bosonic spin-modes are affected by short-ranged processes. The presence of end-states (a feature which is absent in carbon armchair nanotubes~\cite{Mayrhofer2008}), however, smears out the energies within all plateaus (brown diamonds):
%
%
%
%
It induces a \emph{mixing between excited states and groundstates of same total charge and spin}, which widely lifts the degeneracy between the various states. The inset of Fig. \ref{spectrum}, e.g., shows that among eight formerly degenerate groundstates, two get lowered and two get raised by a certain energy under the influence of the end-bulk interaction. We will come across this in more detail during the following analyis. 
\section{Impact on transport}
\begin{figure}
\begin{center}\includegraphics[width=0.99\columnwidth]{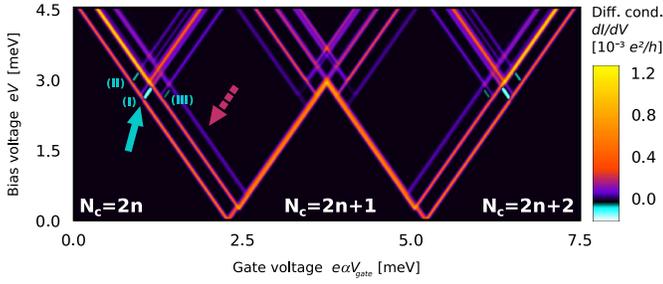}\end{center}
\caption{\label{NDC-plot}Differential conductance of an ACN-SET.
A triplet of states is split by the end-bulk interaction (dashed red arrows).
The green solid arrows point towards regions where negative differential conductance (NDC) is observable. We chose a temperature $T\!=\!116\,$mK and tunneling coupling to the leads $\hbar\Gamma_L\!=\!\hbar\Gamma_R\!=\!10^{-3}\,$meV. All other parameters are as in Fig. 3.}
\end{figure}

In the remaining of this Letter we show how this
entanglement is revealed in the peculiarities in the stability diagram of an ACN-SET.
In the limit of weak coupling to the leads, we can assume that our total system, see also Fig. \ref{graphene}, is described by the Hamiltonian
\begin{equation*}
H=H_\odot+H_{\leads}+H_T -e\alpha V_{\textsf{gate}}{N}_c,
\end{equation*}
with the ACN-Hamiltonian $H_\odot$ given in Eq. (\ref{eq:H_dot}). Further, $H_{\leads}=\sum_{lq}\sum_{\sigma}(\epsilon_q-\mu_{l})c^\dag_{l\sigma q}c_{l\sigma q}$, with $c_{l\sigma q}$ annihilating an electron in lead $l$ of kinetic energy $\epsilon_q$ and the chemical potential $\mu_l$ differs for the left and right contact by $eV$, with $V$ the applied bias voltage. Next, $H_T=\sum_{l\sigma}\int\!\! d^3 r\left(T_l(\vec{r})\psi^\dag_{\sigma}(\vec{r})\phi_{l\sigma}(\vec{r})+h.c.\right)$ descibes tunneling between ACN and contacts, with tunneling coupling $T_l(\vec{r})$ and $\psi_{\sigma}(\vec{r})$ the ACN bulk electron operator as given in Eq. (\ref{elops}), $\phi_{l\sigma_l}(\vec{r})=\sum_{q}\phi_{lq}(\vec{r})c_{l\sigma_lq}$ the lead electron operator with $\phi_{lq}(\vec{r})$ denoting the wave function of the contacts. Finally, the potential term describes the influence
 of a capacitively applied gate voltage $(0\le \alpha \le 1)$.\\
Due to the condition that the coupling between ACN and the contacts is weak,
we can calculate the stationary current by solving a master equation for the
reduced density matrix to second order in the tunneling coupling. As this is a standard procedure, we
refer to \cite{Mayrhofer2006} for details about the method, and show in Fig. \ref{NDC-plot} numerical results for the differential conductance in the $V$-$V_{\textsf{gate}}$ plane. In the numerical calculations an energy cutoff of $1.9\varepsilon_0$ above the groundstate was used, including any energetically allowed bosonic or fermionic excitation. One can clearly observe a two-fold  electron periodicity, with small/large Coulomb diamonds corresponding to even, $N_c=2n$, and odd, $N_c=2n+1$ electron filling.
A triplet of excitation lines is clearly visible in correspondence of the $2n\to 2n\pm 1$ transition (Fig. \ref{NDC-plot}, dashed red arrow). Moreover, NDC occurs as well, despite we considered a  fully symmetric contact set-up (Fig. \ref{NDC-plot}, solid green arrow). To understand these features, it is necessary to consider the eigenstates of the fully interacting ACN in a minimal low-energy model.

 \section{A minimal set of lowest lying states}
 For the following we neglect  short-range bulk-bulk processes as well as the bosonic excitations,
 as they do not  qualitatively change the features we wish to describe.
 For even filling, $N_c=2n$, we consider those eigenstates
$|{\sigma}_L^{\mbox{\small \edge}},\vec{N},{\sigma}_R^{\mbox{\small \edge}}\rangle:= |{\sigma}_L^{\mbox{\small \edge}},\vec{N},\vec{0},{\sigma}_R^{\mbox{\small \edge}}\rangle$ of Eq. (\ref{eq:H0Vrr_diag}) which have total spin $S=0$, no bosonic excitations and up to one fermionic excitation.
This means $\vec{N}=(n,n)$ or $\vec{N}=(n\pm1,n\mp1)$, $n\in\mathbb{N}$. We
introduce the notation
$(n,n):= \, \uparrow\downarrow$, $(n+1,n-1):= \,\uparrow\uparrow$, $(n-1,n+1):=\downarrow\downarrow$ and get then four possible states,\begin{center}\begin{tabular}{rr}
$|a\rangle:=\left|\uparrow,\uparrow\downarrow,\downarrow\right>$,&
$|b\rangle:= \left|\downarrow ,\uparrow\downarrow,\uparrow\right>$,\\
$|c_+\rangle:=\left|\uparrow,\downarrow\downarrow,\uparrow\right>$,&
$|c_-\rangle:=\left|\downarrow,\uparrow\uparrow,\downarrow\right>$.
\end{tabular}\end{center}
%
The states $|a\rangle,|b\rangle$ have the groundstate energy $E^{(0)}_{N_c}=E^{(0)}_{2n}$, while the
 excited states $|c_\pm\rangle$ have energy $E^{(f)}_{2n}=E^{(0)}_{2n}+\varepsilon_0$.
The mixing matrix elements, with
$\cceb$ the end-bulk coupling constant, are
$(V_{\mbox {\small e-b}})_{a c_\pm}=(V_{\mbox {\small e-b}})_{c_\pm a}=\cceb$,
$(V_{\mbox {\small e-b}})_{bc_\pm}=(V_{\mbox{\small e-b}})_{c_\pm b}=-\cceb$. Diagonalization yields:
\begin{eqnarray*}
\mathit{Energy}&:\,&\mathit{Eigenstate\ {(not\ normalized)}}\nonumber\\
\xi_{++}\approx E^{(f)}_{2n}&:\,&\frac{2\cceb}{\xi_{-+}}\left(|a\rangle-|b\rangle\right)+(|c_+\rangle+|c_-\rangle)=:|e_2\rangle
\\
E^{(f)}_{2n}&:\,& |c_+\rangle-|c_-\rangle\hspace{3.16cm}=:|e_1\rangle
\\
\xi_{+-}\approx E^{(0)}_{2n} &:\,&\frac{2\cceb}{\xi_{--}}\left(|a\rangle-|b\rangle\right)+\left(|c_+\rangle+|c_-\rangle\right)=:|g_2\rangle
\\
E^{(0)}_{2n}&:\,& |a\rangle+|b\rangle\hspace{3.6cm}=:|g_1\rangle
\end{eqnarray*}
where
$\label{xi}\xi_{\alpha\alpha'}=\frac{1}{2}(E^{(f)}_{2n}+\alpha E^{(0)}_{2n}+\alpha'\sqrt{\varepsilon_0^2+16 \cceb^2})$.\\ %
In total, the interaction has hardly lifted
the degeneracies between the various states.
However, symmetric and antisymmetric combinations of states $|a\rangle,\,|b\rangle$
 and $|c_+\rangle,\,|c_-\rangle$ arise.
The importance of this mixing becomes obvious when we look now at the states for the odd fillings. As we then necessarily  have an unpaired spin, it is sufficient to consider merely the groundstates, i.e., $\vec{N}=(n\pm1,n)$ with energy $E^{(0)}_{2n+1}$ and total spin $S= \hbar/2$.
%
\begin{figure}
\begin{center}\includegraphics[width=0.86\columnwidth]{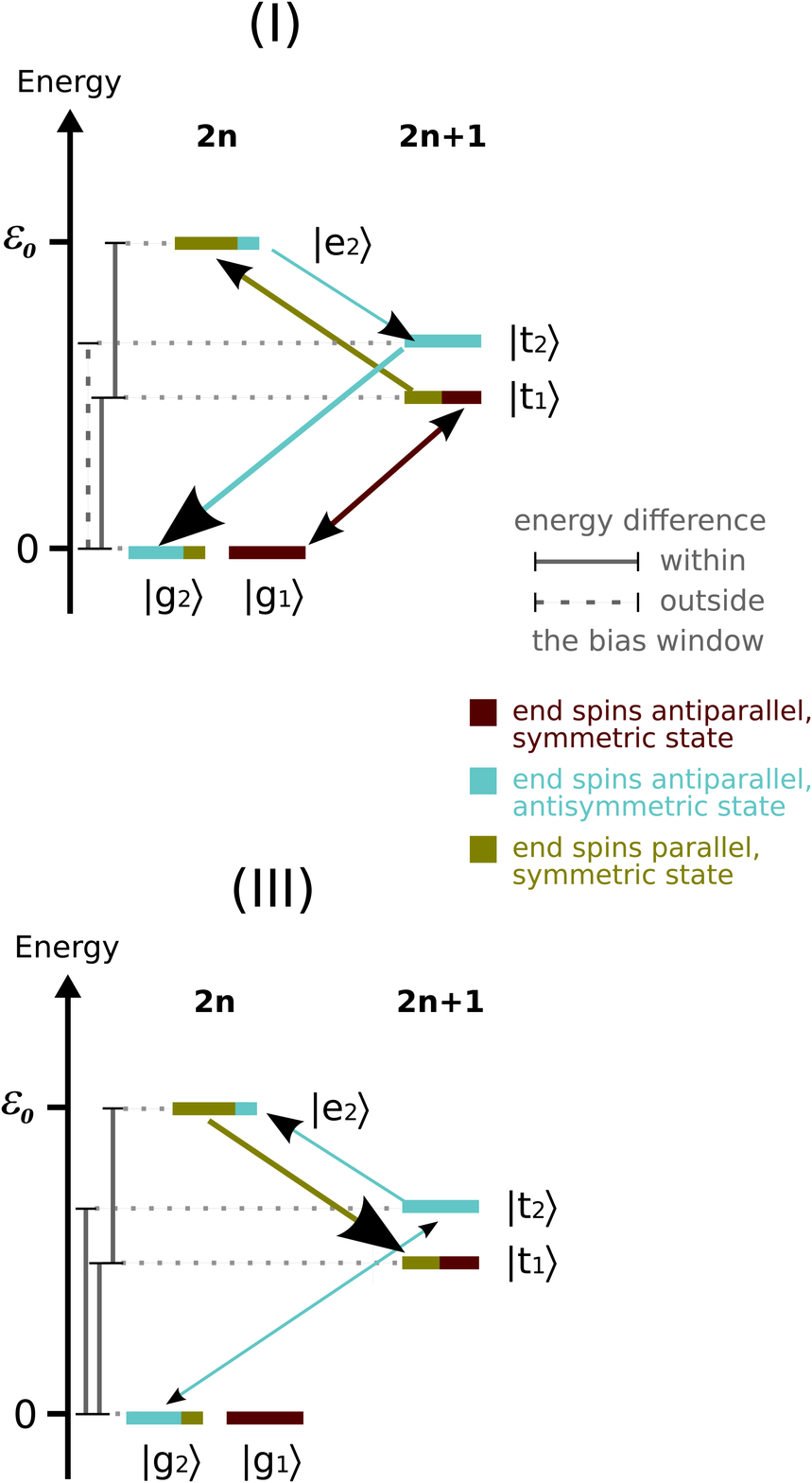}\end{center}
\caption{Schematic explaining the mechanisms causing the NDC features (I) and (III) in Fig. \ref{NDC-plot}. Only states and transitions relevant for the NDCs are drawn. The crucial transition is marked by a big arrow head. (I) Opening of the channel \teins$\to$\ezwei{} leads to a decay into the trapping state \gzwei, depleting the transport channel \geins$\leftrightarrow$\teins. (III) Opening of the channel \tzwei$\to$\ezwei{} depletes the transport channel \gzwei$\leftrightarrow$\tzwei.\label{scheme}}
\end{figure}
We introduce the notation,  $(n+1,n):=\,\uparrow, (n-1,n):=\,\downarrow$ and find the
 six states\begin{center}\begin{tabular}{ll}
$|a_\uparrow\rangle:=\left|\uparrow,\uparrow,\downarrow\right> $,&
$|a_\uparrow\rangle:=\left|\uparrow,\downarrow,\downarrow\right>$\\
$|b_\uparrow\rangle:=\left|\downarrow,\uparrow,\uparrow\right>$,&
$|b_\downarrow\rangle:=\left|\downarrow,\downarrow,\uparrow\right>$,\\
$|c_\uparrow\rangle:=\left|\uparrow,\downarrow,\uparrow\right>$,&
$|c_\downarrow\rangle:=\left|\downarrow,\uparrow,\downarrow\right>$.
\end{tabular}\end{center}
%
The mixing matrix elements read ($\star \in \{\uparrow, \downarrow\}$):\\
  $(V_{\mbox {\small e-b}})_{a_\star c_\star}=(V_{\mbox {\small e-b}})_{c_\star a_\star}=(V_{\mbox {\small e-b}})_{b_\star c_\star}=(V_{\mbox{\small e-b}})_{c_\star b_\star}=-\cceb$. Diagonalization yields:
\begin{subequations}
\begin{eqnarray*}
\mathit{Energy}&:\,&\mathit{Eigenstate\ {(not\ normalized)}}\nonumber\\
E^{(0)}_{2n+1}+\sqrt{2}\cceb&:\,& |t_3\rangle:=|a_\star\rangle+|b_\star\rangle-\sqrt{2}c_\star\hspace{0.2cm}=:|t_3\rangle,
\\
E^{(0)}_{2n+1}&:\,&|a_\star\rangle-|b_\star\rangle\hspace{2.49cm}=:|t_2\rangle,
\\
E^{(0)}_{2n+1}-\sqrt{2}\cceb&:\,& |a_\star\rangle+|b_\star\rangle+\sqrt{2}|c_\star\rangle\hspace{1.04cm}=:|t_1\rangle.
\end{eqnarray*}
\end{subequations}
\section{The excitation line triple}
Compared to the even fillings, the interaction induced lifting of the formerly degenerate $2n+1$ states is much more pronounced and  seizable in the stability diagram of Fig. \ref{NDC-plot} in form of the triple of three parallel lines the dashed red arrow points to. The splitting has the expected value of $\sqrt{2}\cceb$. In detail, the lines mark transitions from the $2n$ groundstates \geins{} and \gzwei{} to the $2n+1$ states \teins, \tzwei{} and \tdrei. Hereby, the antisymmetric state \tzwei{}, associated to the second line of the triple, is special, because it is the only one strongly connected to the $2n$ state \gzwei. 
The first line of the triple is the \geins$\to$\teins{} groundstate transition line.
\begin{figure}
\begin{center}
\includegraphics[width=0.65\columnwidth]{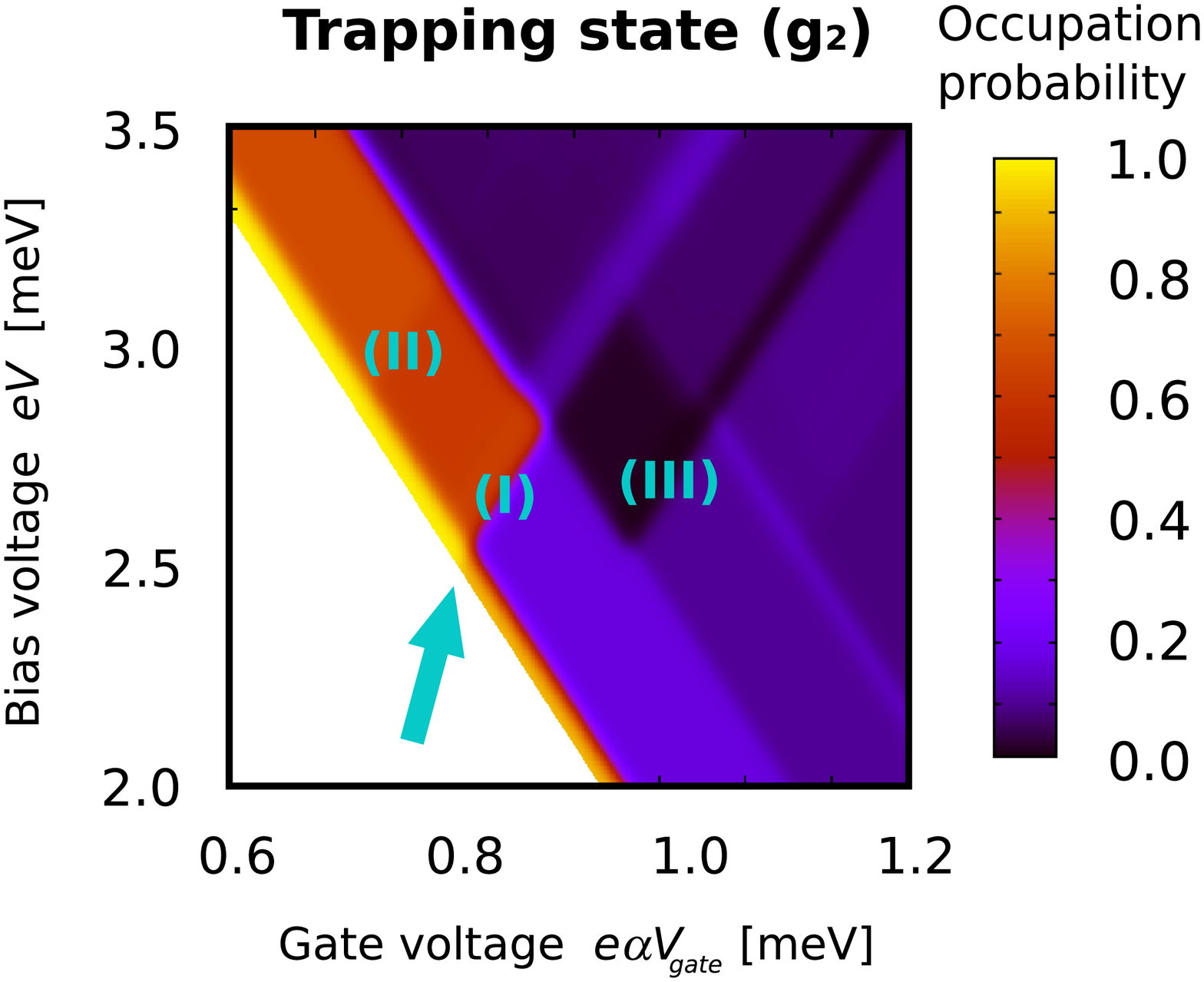}
\end{center}
\caption{Occupation probability of the trapping state \gzwei{} around the region exhibiting various NDC features. Their positions are marked according to Fig. \ref{NDC-plot} with labels (I)-(III). Notice that no numerically stable data can be obtained inside the Coulomb diamond.\label{occ-g1}}
\end{figure}
\section{The NDC mechanism}The NDC (I) highlighted by the solid green arrow marks the opening of the $2n+1\to 2n$ back-transition channel \teins$\to$\ezwei{}. The situation is sketched in Fig. \ref{scheme}. Once \ezwei{} gets populated, from this excited $2n$ states the system can decay into any of the lowest lying $2n+1$ states, and in particular there is a chance to populate the antisymmetric state \tzwei{}. This state is strongly connected to the $2n$ groundstate \gzwei{}, which contains a large contribution of the antisymmetric combination $|a\rangle-|b\rangle$. But in the region where the NDC occurs, the forward channel \gzwei$\to$\tzwei{} is not yet within the bias window such that \gzwei{} serves as a trapping state. Fig. \ref{occ-g1} confirms this explanation: the population of the state \gzwei{} is strongly enhanced in the concerned region where the back-transition \teins$\to$\ezwei{} can take place, while the forward transition \gzwei$\to$\tzwei{} is still forbidden.\\
In a completely analog way, just involving instead of \ezwei{} an excited $2n$ state with total spin $\hbar$ (not listed before), NDC (II) arises.\\
The origin of NDC (III) is of different nature. It belongs to the back-transition \tzwei$\to$\ezwei{}, which is a weak channel because \tzwei{} is a purely antisymmetric state, while the antisymmetric contribution in \ezwei{} is rather small. From time to time, nevertheless the transition will take place, and once it happens the system is unlikely to fall back to \tzwei, but will rather change to a symmetric $2n+1$ state. 
Thus the state \tzwei{} is depleted, and with it the transport channel \tzwei$\leftrightarrow$\gzwei, which leads to NDC. The statement can also be verified from the plot of the occupation probability for \gzwei, Fig. \ref{occ-g1}: a pronounced dark region of decreased population follows upon the NDC transition.
\section{Summary}
In conclusion, we focussed on  small-width ACNs, and showed that the low energy properties are
dominated by entangled bulk-end states. One major consequence is that the bulk spin is not conserved and
that the symmetry of the entangled states  generates trapping states and hence negative differential conductance.\\


We acknowledge the support of the DFG under the programs SFB 689 and GRK 638.

\end{document}